\documentstyle[aps,multicol]{revtex}

\newcommand{\beq}{\begin{equation}}
\newcommand{\eeq}{\end{equation}}
\newcommand{\beqa}{\begin{eqnarray}}
\newcommand{\eeqa}{\end{eqnarray}}

\def\half{\frac{1}{2}}
\def\<{\langle}
\def\>{\rangle}

\begin{document}
\title{Spin flips and quantum information for anti-parallel spins}

\author
{N. Gisin$^a$ and S. Popescu$^{bc}$\\
{$^a$\protect\small\em Group of Applied Physics, University of Geneva, 1211
Geneva 4, Switzerland}\\
{$^b$\protect\small\em Isaac Newton Institute, University of Cambridge, 20 
Clarkson Road, Cambridge, CB3 0EH, UK}\\
{$^c$\protect\small\em BRIMS Hewlett Packard Labs, Bristol, UK}
}
\date{\today}

\maketitle

\begin{abstract}
We consider two different ways to encode quantum information, by parallel or 
anti-parallel pairs of spins. We find that there is more information in the 
anti-parallel ones. This purely quantum mechanical effect is due to 
entanglement, not of the states but occuring in the course of the measuring 
process. We also introduce a range of quantum information processing machines, 
such as spin-flip and anti-cloning.
\end{abstract}

\begin{multicols}{2}

\section{Introduction}\label{int}
Quantum information differs from classical information because it obeys the
superposition principle and because it can be entangled. The huge potential of 
quantum information processing has renewed the interest
in the foundations of two of the major scientific theories of the twentieth
century:
information theory and quantum mechanics \cite{PhysWorld98}.

Despite the very intensive recent work on quantum information, surprising
effects are continuously being discovered.  Here we describe yet another 
surprise. In 
a nutshell, we enquire whether quantum information is better stored by two 
parallel spins or two anti-parallel ones?

In more detail, our paper is centered around the
following problem of quantum communication.  Suppose Alice wants to communicate
Bob a space direction $\vec n$.  She may do that by one of the following two
strategies.  In the first case, Alice sends Bob two spin 1/2 particles polarized
along $\vec n$, i.e.  $|\vec n, \vec n>$.  When Bob receives the spins, he
performs some measurement on them and then guesses a direction $\vec n_{g}$
which has to be as close as possible to the true direction $\vec n$.  The second
strategy is almost identical to the first, with the difference that Alice sends
$|\vec n, -\vec n>$., i.e.  the first spin is polarized along $\vec n$ but the
second one is polarized in the {\it opposite} direction.  The question is
whether these two strategies are equally good or, if not, which is better.

To put things in a better perspective, consider first a simpler problem.
Suppose Alice wants to communicate Bob a space direction $\vec n$ and she may do
that by one of the following two strategies.  In the first case, Alice sends Bob
a single spin 1/2 particle polarized along $\vec n$, i.e.  $|\vec n>$.  The 
second
strategy is identical to the first, with the difference that when Alice wants to
communicate Bob the direction $n$ she sends him a single spin 1/2 particle
polarized in the opposite direction, i.e.  $|-\vec n>$.  Which of these two
strategies is better?

If the particles would be classical spins then, obviously, both strategies would 
be equally good, as an arrow defines equally well both the direction in which it 
points and the opposite direction. Is the quantum situation the same?
 
First, we should note that in general, by sending a single spin 1/2 particle, 
Alice cannot communicate to Bob the direction $\vec n$  with absolute precision.
Nevertheless, it is still obviously true that the two strategies are equally 
good. Indeed, all Bob has to do in the second case is to perform exactly the 
same measurements as he would do in the first case, only that when his results 
are such that in the first case he would guess $\vec n_g$, in the second case he 
guesses $-\vec n_g$. 

We are thus tempted to conjecture that:

\noindent
{\bf Conjecture:} Similar to the classical case, for the 
purpose of defining a direction $\vec n$, a quantum mechanical spin polarized 
along $\vec n$ is as good as a spin polarized in the opposite direction. In 
particular, the two two-spin states $|\vec n, \vec n>$ and $|\vec n, -\vec n>$ 
are equally good. 

Surprisingly however, as we'll show here, this conjecture is not true.

As we will show, the main reason behind this effect is, once more, entanglement. 
Here entanglement does not refer to the two spins - whether parallel or 
anti-parallel they are always in a direct product state - but to the 
eigenvectors of the optimal measurement. (Indeed, as Massar and Popescu 
demonstrated \cite{MassarPopescu1995} the optimal measurement on 
parallel spins requires entanglement.) 

This result has also led us to many new questions. For example we were led to 
consider a {\it universal quantum spin flip (UQSF) machine} a machine which
flips an unknown spin as well as possible, and an {\it anti-cloning} machine, 
i.e.
a machine which takes as input N parallel spins, polarized in an unknown
direction, and generates some supplementary spins polarized in the opposite
direction.  We also point out the relation between spin flip
and partial transpose.

\section{Optimal fidelity for parallel and anti-parallel 
spins}\label{optfidelity}

In the previous section we have presented our main problem as a quantum 
communication problem. We can present it also in a different way, which brings 
it closer a well-known problem. Indeed, we can completely dispense with Alice, 
and 
consider that there is a source which emits pairs of parallel (or anti-parallel) 
spins, and Bob's task is to identify the state as well as possible. 

For concreteness, let us define Bob's measure of success as the fidelity,

\beq
F=\int d\vec n \sum_g P(g|\vec n){{1+\vec n\vec n_g}\over 2}
\eeq
where $\vec n\vec n_g$ is the scalar product in between the true and the 
guessed directions, the integral is over the different directions $\vec n$ and 
$d \vec n$ represents the {\it a priori} probability that a state associated to 
the 
direction 
$\vec n$, i.e. $|\vec n, \vec n>$ or $|\vec n, -\vec n>$ respectively, is 
emitted by the source; $P(g|\vec n)$ is the probability of guessing $\vec n_g$
when the true direction is $\vec n$. In other 
words, for each trial Bob gets a score which is a (linear) function of the 
scalar product 
between the true and the guessed direction, and the final score is the 
average of the individual scores.

When the different directions $\vec n$ are randomly and uniformly distributed 
over the unit sphere, an optimal measurement for pairs of parallel spins
$\psi=|\vec n, \vec n>$ has been found by Massar and Popescu 
\cite{MassarPopescu1995}.  Bob has to measure an operator A whose eigenvectors 
$\phi_j$, $j=1...4$ are
\beq
|\phi_j>=\frac{\sqrt{3}}{2}|\vec n_j, \vec n_j> + \frac{1}{2}|\psi^->
\eeq
where $|\psi^->$ denotes the singlet state and the Bloch vectors $\vec n_j$
point to the
4 vertices of the tetrahedron:
\beqa
\vec n_1&=&(0,0,1)  \nonumber \hspace{1cm}
\vec n_2=(\frac{\sqrt{8}}{3},0,-\frac{1}{3})  \nonumber \\
\vec n_3&=&(\frac{-\sqrt{2}}{3},\sqrt{\frac{2}{3}},-\frac{1}{3})  \hspace{1cm}
\vec n_4=(\frac{-\sqrt{2}}{3},-\sqrt{\frac{2}{3}},-\frac{1}{3})
\eeqa
[The phases used in the definition of $|\vec n_j>$ are such that the 4 states 
$\phi_j$ are mutually orthogonal].
The exact values of the eigenvalues corresponding to the above eigenvectors are 
irrelevant; all that is important is that they are different from each other, so 
that each eigenvector can be unambiguously associated to a different outcome of 
the measurement. 
If the measurement results corresponds to $\phi_j$, then the guessed direction 
is $\vec n_j$. The corresponding optimal fidelity is 3/4 
\cite{MassarPopescu1995}.

A related case is when the directions $\vec n$ are {\it a priori} on the 
vertices of 
the 
tetrahedron, with equal probability 1/4. Then the above measurement provides
a fidelity of
5/6$\approx 0.833$, conjectured to be optimal.

Let us now consider pairs of anti-parallel spins, $|\psi>=|\vec
n,-\vec n>$,
and the measurement whose  eigenstates are
\beqa
\theta_j=\alpha|\vec n_j,-\vec n_j> - \beta\sum_{k\neq j} |\vec n_k,-\vec n_k>
\eeqa          
with $\alpha=\frac{13}{6\sqrt{6}-2\sqrt{2}}\approx 1.095$ and 
$\beta=\frac{5-2\sqrt{3}}{6\sqrt{6}-2\sqrt{2}}\approx 0.129$.  
The corresponding fidelity for uniformly distributed $\vec n$ is
$F=\frac{5\sqrt{3}+33}{3(3\sqrt{3}-1)^2}\approx 0.789$; and for $\vec n$
lying on the tetrahedron $F=\frac{2\sqrt{3}+47}{3(3\sqrt{3}-1)^2}\approx 0.955$.
In both cases the fidelity obtained for pairs of anti-parallel spins is
larger than
for pairs of parallel spins!

\section{Spin flips}\label{flips}

As we have seen in the previous section, parallel and anti-parallel spins are 
not equivalent. Let us try to understand why.

That there could be any difference between communicating a direction by two 
parallel spins or two anti-parallel spins seems, at first sight, extremely 
surprising. After all, by simply flipping one of the spins we could change one 
case into the other. For example, if Bob knows that Alice indicates the 
direction by two anti-parallel spins he only has to flip the second spin and 
then 
apply all the measurements as in the case in which Alice send from the beginning 
parallel spins. Thus, apparently, the two methods are bound to be equally good. 

The problem is that one cannot flip a spin of unknown polarization. Indeed, it 
is easy to see that the flip operator V defined as
\beq
V|\vec n>= |-\vec n>
\eeq
is not unitary but anti-unitary. Thus there is no physical operation which could 
implement such a transformation.

But then a couple of question arise.  First, why is it still the case that a
single spin polarized along $\vec n$ defines the direction as well as a single
spin polarized in the opposite direction?

What happens is that although Bob cannot implement an {\it active} 
transformation, i.e. cannot flip the spin, he can implement a {\it passive} 
transformation, that is, he can flip his measuring devices. Indeed, there is no 
problem for Bob in flipping all his Stern-Gerlach apparatuses, or, even simpler 
than 
that, to merely rename the outputs of each Stern-Gerlach ``up"$\rightarrow$ down 
and ``down"$\rightarrow$``up". 

But given the above, why can't Bob solve the problem of two spins in the same 
way, 
just by 
performing a passive transformation on the apparatuses used to measure the 
second 
spin?!? The problem is the entanglement. Indeed, if the optimal strategy for 
finding the polarization direction  would involve separate measurements on the 
two spins then two parallel spins would be equivalent to two anti-parallel 
spins. (This would be true even if which measurement is to be performed on the 
second spin depends on the result of the measurement on the first spin.) But, as 
shown in the previous section, the optimal measurement is {\it not} a 
measurement performed separately on the two spins but a measurement which acts 
on both spins simultaneously, that is, the measurement of an operators whose 
eigenstates are entangled states of the two spins. For such a measurement there 
is no way of associating different parts of the measuring device with the 
different spins, and thus there is no way to make a passive flip associated to 
the second spin. Consequently there is no way, neither active nor passive to 
implement an equivalence between the parallel and anti-parallel spin cases.

This result illustrates once more that entanglement can produce results
``classically
impossible", similar to Bell inequality and to non-locality without
entanglement \cite{MassarPopescu1995,Bennett}.

\section{Spin flips and the partial transpose of bipartite density 
matrices}\label{transpose}

We have claimed in the previous section that when we perform a measurement of an 
operator whose eigenstates are entangled states of the two spins, there is no 
way 
of making a passive flip associated with the second spin. We would like to 
comment 
in more detail about this point. 

Physically it is clear that in the case of a measuring device corresponding to 
an operator whose eigenstates are entangled states of the two spins, we cannot 
identify one part of the apparatus as acting solely on one spin and another part 
of the apparatus as acting on the second spin. Thus we cannot simply isolate a 
part of the measuring device and rename its outcomes. But perhaps one could make 
such a passive transformation at {\it mathematical} level, that is, in the 
mathematical description of the operator associated to the measurement and then 
physically construct an apparatus which corresponds to the new operator.

The optimal measurement on two parallel spins is described by a nondegenerate 
operator whose eigenstates $|\phi_j>$ are given by (2) and (3). It is convenient 
to consider the projectors $P^j =|\phi_j><\phi_j|$ associated with the 
eigenstates. As is well-known, any unit-trace hermitian operator, and in 
particular any  
projectors, can be written as

\beq
P^j={1\over4}(I +\vec\alpha^j\vec\sigma^{(1)}+ 
\vec\beta^j\vec\sigma^{(2)}+R_{k,l}^j\sigma^{(1)}_k\sigma^{(2)}_l).
\label{Pj}
\eeq
with some appropriate coefficients $\vec\alpha^j$, $\vec\beta^j$ and 
$R_{k,l}^j$. (The upper indexes on the spin operators mean ``particle 1" or 
``2"). 
Why then couldn't we simply make the passive spin flip by considering a 
measurement  described by the projectors
\beq
\tilde P^j={1\over4}(I +\vec\alpha^j\vec\sigma^{(1)}- 
\vec\beta^j\vec\sigma^{(2)}-R_{k,l}^j\sigma^{(1)}_k\sigma^{(2)}_l).
\label{tPj}
\eeq
obtained by the flip of the operators associated second spin, 
$\vec\sigma^{(2)}\rightarrow -\vec\sigma^{(2)}$? The reason is that the 
transformed operators $\tilde P^j$ are no longer projectors! Indeed, each 
projection 
operator $P^j$ could also be viewed as a density matrix 
$\rho^j=P^j=|\phi_j><\phi_j|$. The passive spin flip (\ref{Pj})$\rightarrow$
(\ref{tPj}) is nothing more that the 
partial 
transpose of the density matrices $\rho^j$ with respect to the second spin. But 
each density matrix $\rho^j$ is {\it non-separable} (because they describe the 
entangled state $|\Phi>$). But according to the well-known result of the 
Horodeckis \cite{PartilaTPeres,horo} the partial transpose of a non-separable 
density matrix 
of two spin 1/2 particles has a negative eigenvalue and thus 
it cannot represent a projector anymore. Obviously however, if the optimal 
measurement would have consisted of independent measurements on the two spins, 
each projector would have been a direct product density matrix and the spin flip 
would have transformed them into new projectors, and thus led to a valid new 
measurement.

\section{Spin flips, entropy and the global structure of the set of 
states}\label{entropy}

There is yet another surprise in the fact that anti-parallel spins can be better 
distinguished than parallel ones. Consider the two sets of states, that of 
parallel spins and that of anti-parallel spins. The distance in between any two 
states in the first set is equal to the distance in between the corresponding 
pair of states in the second set. That is, 
\beq
|<\vec n, \vec n|\vec m, \vec m>|^2=|<\vec n, -\vec n|\vec m, -\vec m>|^2
\eeq
Nevertheless, {\it as a whole}, the anti-parallel spin states are farer apart
than the parallel ones!  Indeed, the anti-parallel spin states span the entire
4-dimensional Hilbert space of the two spin $\half$, while the parallel spin
states span only the 3-dimensional subspace of symmetric states.
This is similar to a 3 spin example discovered by R. Jozsa and J. 
Schlienz\cite{Jozsa}.

\section{The universal quantum spin-flip and anti-cloning machines}
As we have already noted in section III, a perfect universal quantum spin-flip 
machine i.e. a machine which would reverse any spin $\half$
state $|\vec n>\rightarrow|-\vec n>$ is impossible -  it would require an 
anti-unitary transformation. However, following the lesson of the cloning 
machine, \cite{Buzek}, let us ask how well could one approximate such a machine. 

Analog to an optimal universal cloning machine, let us define an optimal {\it 
universal quantum spin-flip machine (UQSF)}. By definition, a UQSF is a machine 
which 
acting on a spin 1/2 particle implements the transformation

\beq
|\vec n>\rightarrow \rho(\vec n)
\label{spinflip}
\eeq
such that $\rho(\vec n)$ is as close as possible to $|-\vec n>$. For 
concreteness, we define ``as close as possible" to mean ``according to the usual 
fidelity" $F=\int d\vec n<-\vec n|\rho(\vec n)|-\vec n>$.  Furthermore, to be 
``universal" 
we 
require that the fidelity is independent of the initial polarization of the 
spin, that is, that all states are flipped equally well. (Obviously, in order to 
be able to implement the transformation (\ref{spinflip}) which is non-unitary, 
the UQSF 
machine is allowed to entangle the spin with an ancilla.)

Following the technique of \cite{6state}, developed for optimal eavesdropping in 
the
six state protocol of quantum cryptography, one finds that the fidelity of the
optimal quantum spin-flip machine is of 2/3 (which appears as the maximal
disturbance Eve can introduce in the quantum channel).  

One simple way to implement this optimal spin flip consists in first measuring
the spin in an arbitrary direction, then produce a spin pointing in the
direction opposite to the measurement result.

Surprisingly enough, although the original goal was to flip a single spin (the
input spin) the optimal UQSF machine can produce additional flipped spins at no
extra cost! This follows from the fact that the optimal UQSF provides {\it 
classical} information which then can be used to prepare as many flipped spins 
as we want. This result is surprising because one is tempted to imagine that if 
we only want to flip a single spin we could do it with much better fidelity if 
we don't 
attempt to extract classical information from it. At least this is the lesson 
of many other quantum information processing procedures, such as cloning, 
teleportation, data compression, quantum computation etc. In all these cases 
quantum information can be processed with much better results if we keep it 
all the time in quantum form rather than extracting some classical information 
from it and processing this classical information. The deep reason why
spin flipping is essentially a classical operation is an interesting but 
yet open question.

One can also consider other interesting machines. For example machines that 
take as input 2 parallel spins $|\vec n,\vec n>$ and the output is as close as 
possible either to $|\vec n,-\vec n>$, or to $|-\vec n>$, or to 
$|\vec n,\vec n,-\vec n>$. Another interesting open question is which 
operation can be produced with 
higher fidelity: from two parallel spins to anti-parallel ones, or vice versa.

\section{Spin-flips and quantum optics}
Entanglement is closely connected to the mathematics of partial transpose
$\rho^T_{ij,kl}
=\rho_{il,kj}$: a 2-spin $\half$ (mixed) state is separable if and only if its
partial transpose has non-negative eigenvalues
\cite{PartilaTPeres,horo}. 
Interestingly, partial transposes can be seen as a representation of spin
flips. 
Indeed, the partial transpose of a product operator reads
\beq
(a_0+\vec a\vec\sigma)\otimes(b_0+\vec b\vec\sigma)\rightarrow(a_0+\vec
a\vec\sigma)\otimes(b_0+\vec b\vec\sigma-2b_z\sigma_z)
\eeq
(where the $\sigma_k$ are the usual Pauli matrices)
hence, a partial transpose is a reflection of the second spin through the
x-z plane
(the plane depends on the basis, as partial transpose is basis-dependent).
Note that
this is a practical way of representing the polarization of a photon
reflected by a
mirror: the upper and lower hemisphere of the Poincar\'e sphere are exchanged,
corresponding to the change of right handed and left handed elliptic
polarization states.
To complete the connection with spin flips, add after the reflection a $\pi$
rotation
around the axes orthogonal to the reflection plane (like the Faraday rotator in 
Faraday mirrors \cite{FM}), this flips the second spin. 
Now, the proof that perfect UQSF machines do not exist can be reformulated:
a perfect UQSF machine would turn entangled states into states with negative
eigenvalues!
Physically, the use of a mirror acting only on the second photon is of
course still possible,
but one must note that mirrors change right handed reference frames into
left handed ones.
This is acceptable as long as one can describe the two photons separately,
but leads to erroneous
predictions if applied to entangled photons.

\section{Conclusion}\label{conc}
We have proved that there is more information about a space direction $\vec n$ 
in a pair of antiparallel spins $|\vec n,-\vec n>$ than in a pair of
parallel spins $|\vec n,\vec n>$. This demonstrates again the role played by 
entanglement in quantum information processing: not a source of paradoxes, but a 
means of performing tasks which are impossible classically. It also draws the 
attention to 
the global structure of the state space of combined systems. Related questions 
concern the optimal quantum spin-flip machine and the optimal quantum machine 
that turns parallel to anti-parralel pair of spins.

\small
\section{Acknowledgments}
This work was partially supported by the Swiss National Science Foundation
and by
the European TMR Network ``The Physics of Quantum Information" through the Swiss
OFES.
After completion of this work, the very interesting article by V. Buzek, M. 
Hillery and R. 
Werner appeared on the quant-ph \cite{QUNOTG} introducing the universal quantum 
NOT gate which is our quantum spin-flip machine. Finally, we like to thank Lajos 
Diosi for helpful discussions on spin-flips.

\normalsize

\end{multicols}
\end{document}